\documentclass[aps,prb,twocolumn]{revtex4} 

\usepackage{graphicx}
\usepackage{dcolumn}
\usepackage{bm}

\newcommand{\comment}[1]{}

\begin{document}
\renewcommand{\theequation}{\arabic{section}.\arabic{equation}}

\title{Design of Chemotaxis Devices Using Nano-Motors}


\author{Phil Attard}

\date{3 September, 2012, phil.attard1@gmail.com, arXiv:cond-mat}

\begin{abstract}
Several designs for micro-devices for chemotaxis
based on nano-motors are proposed.
The nano- or micro-motors are the conventional Janus rods or spheres
that are powered by the catalytic reaction of fuels such as
hydrogen peroxide.
It is shown how these can be linked to make a device
that can follow a concentration gradient of the fuel.
The feasibility of assembling the devices using micromanipulation
or metallic deposition is discussed.
A possible design principle is suggested for a device
that follows the concentration gradient of an analyte other than the fuel.
\end{abstract}


\maketitle

                \section{Introduction}

Chemotaxis is the movement of biological organisms
in response to a chemical concentration gradient.
Three things are required  for the phenomenon:
an ability to sense the gradient,
a means of steering or orientation,
and a means of propulsion.
It is a challenge
to modern micro-mechanical techniques
to manufacture an artificial chemotaxis device
on similar scales to those that occur in real organisms.
One might imagine in the long term
that such devices might play a role in targeted drug delivery
or in chemical detection and localization.
A first step on the long journey to such possible applications
is the conceptual design of a micron sized chemotaxis device.

In recent years micro-motors
have been manufactured in the form of Janus spheres or rods
that differentially catalyze reactions at their surface.
The solvent or solute provide the fuel,
and the differential reaction is the basis of the propulsion mechanism.
A common example is the catalytic decomposition of hydrogen peroxide,
in which case
both non-conducting (e.g.\ Pt/SiO$_2$, Pt/TiO$_2$)
and conducting (e.g.\ Au/Pt) combinations
have been used.\cite{Paxton05,Kline05,Ozin05,Howse07,Burdick08,Gibbs09}
In the conducting case,
the propulsion mechanism
is thought to be electrokinetic.\cite{Moran11}

One limitation of the artificial micro-motors
that have been manufactured to date
is that they have no method of sensing
a concentration gradient or of aligning themselves with it.
Although the propulsion force and hence the speed
of the differential catalytic motors
is proportional to the local concentration of fuel,
this in itself does not provide directed motion
because random Brownian impulses rapidly disorient the motor
and there is no restoring mechanism to realign it in the preferred direction.
Hence the motor follows a random walk through the solvent,
with speed proportional to the local concentration of fuel,
but with the velocity on average zero.
Directional control has been achieved
by using an external magnetic field and incorporating a ferromagnetic segment
in the micro-motor,\cite{Kline05,Burdick08}
but obviously
as a model of chemotaxis it would be preferable
if the steering mechanism
could be based on the chemical concentration gradient itself.

The present note proposes some relatively simple designs
that link differential catalytic micro-motors
to achieve directional motion due to a chemical concentration gradient.
The following section gives two designs for such chemotaxis micro-devices.
In the third section,
the feasibility of manufacture and other practical considerations
are discussed.

                \section{Chemotaxis Micro-Devices}

\begin{figure}[b]
\centerline{ \resizebox{2.5cm}{!}{ \includegraphics*{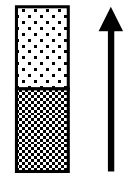}
} }
\caption{\label{mm-element}
Micro-motor element.
The solid arrow indicates the direction of motion.}
\end{figure}

Figure \ref{mm-element} is a sketch of a micro-motor.
The convention used in this note
is that the force on the motor is from the dark end to the light end,
and hence the motion tends in the direction signified by the arrow.
To be concrete,
if the micro-motor was a Pt/Au Janus rod
and the fuel was H$_2$O$_2$,
then gold would be dark and platinum light.
As mentioned in the introduction,
the micro-motor has no preferred orientation
and so it always travels in whatever direction its axis happens
to be pointing irrespective  of any concentration gradient.
Random fluctuations frequently reorient the motor.
The propulsive force is proportional to the local concentration.
In practice the motor reaches a terminal speed
in which the drag force is equal and opposite to the propulsive force.
Typical dimensions of the motor are 2$\,\mu$m long
and 300$\,$nm in diameter,
and speeds can exceed 200$\,\mu$m$\,$s$^{-1}$.\cite{Moran11}

\begin{figure}[t!]
\centerline{ \resizebox{6.cm}{!}{ \includegraphics*{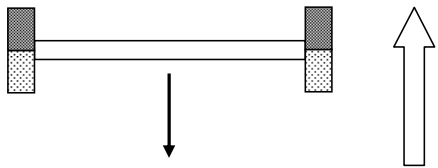}
} }
\caption{\label{mm-ataxis}
Device for stable negative chemotaxis.
The device consists of two motor elements attached to
an inactive bar.
The device is drawn in the stable orientation,
with the solid arrow indicating the direction of motion,
and the block arrow indicating the concentration gradient.}
\end{figure}

\begin{figure}[t!] \vspace{.3cm}
\centerline{ \resizebox{6.cm}{!}{ \includegraphics*{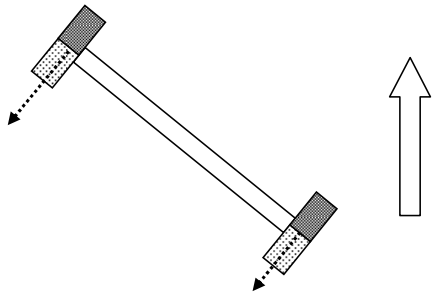}
} }
\caption{\label{mm-ataxis-rotate}
Rotated device.
The forces on the motors (dotted arrows, length proportional to magnitude)
sum to give a SW force
and a counter-clockwise torque.}
\end{figure}

The basic concept for the chemotaxis devices
is to use micro-motors to create a torque
in the presence of a concentration gradient.
The simplest such device is sketched in Fig.~\ref{mm-ataxis}.
(Practical issues regarding the manufacture and performance of the
chemotaxis devices will be discussed in the following section.)
The device consists of two parallel micro-motors,
oriented perpendicular to a neutral lever arm
and rigidly attached at the ends.
It makes no difference whether the attached motors are Janus cylinders
or spheres.
In the orientation shown,
the device moves against the concentration gradient,
(i.e.\ from high concentrations to low),
which is called negative chemotaxis.
The reason for this particular arrangement is that it is stable
to random fluctuations.
For example, in Fig.~\ref{mm-ataxis-rotate}
the device is rotated clockwise 45$^\circ$.
Both motors create a propulsive force in the SW direction,
but, due to the concentration gradient,
the left motor experiences a higher concentration
and hence creates a larger force.
The consequent net turning moment on the couple
acts in the counter-clockwise direction,
which tends to restore the device to the original orientation.

For simplicity, in Fig.~\ref{mm-ataxis}
the device is sketched in the plane of the page,
and the discussion focussed on angular fluctuations within the plane of the page.
In reality one has to be concerned with the orientation of the device
in  three dimensional space.
The device as drawn in Fig.~\ref{mm-ataxis} would experience random
twists about the long axis that would provide a stochastic contribution
to its motion superimposed upon the deterministic negative chemotaxis.
For full stability,
one ought attach to the middle of the device in Fig.~\ref{mm-ataxis}
an identical device oriented perpendicular to the page
(all four motors parallel).
Such an X-shaped device is stable with respect to all angular fluctuations.

\begin{figure}[t]
\centerline{ \resizebox{6.cm}{!}{ \includegraphics*{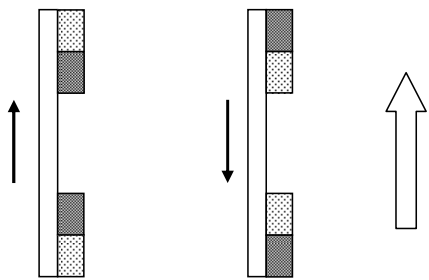}
} }
\caption{\label{mm-steering}
Steering elements.
The direction of motion (solid arrows) is shown
when the element is aligned with the concentration gradient (block arrow).}
\end{figure}

A limitation of the simple device sketched in Fig.~\ref{mm-ataxis}
(and its three dimensional, cross-shaped analogue)
is that it only performs negative chemotaxis.
It is desirable to have the possibility of positive chemotaxis.
This can be accomplished by using steering elements
of the form shown in Fig.~\ref{mm-steering}.
A steering element consists of two antiparallel motors
rigidly attached at the ends of a rod with which they are aligned.
In the orientation shown,
the elements move either with (positive chemotaxis, left element)
or against (negative chemotaxis, right element)
the concentration gradient.
Even though the motors on each element oppose each other,
due to the concentration gradient the motor at the higher concentration
provides the larger propulsive force
and hence the direction of motion.
Neither steering element on its own has any reorientation ability
and so the configuration shown is not stable.

\begin{figure}[t!]
\centerline{ \resizebox{8.cm}{!}{ \includegraphics*{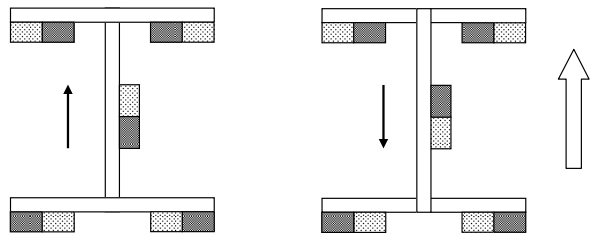}
} }
\caption{\label{mm-stable}
Stable chemotaxis.
The solid arrows indicate the directions of motion,
and the block arrow indicates the concentration gradient.}
\end{figure}

\begin{figure}[t!]
\centerline{ \resizebox{5.cm}{!}{ \includegraphics*{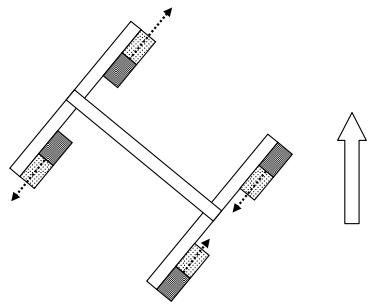}
} }
\caption{\label{mm-stable-rotate}
Rotated device (without motor).
The dotted arrows are the force vectors,
which sum to give a clockwise torque.}
\end{figure}

Joining the two types of steering elements with a rigid rod,
as in Fig.~\ref{mm-stable},
provides the necessary orientational stability.
For example, in Fig.~\ref{mm-stable-rotate}
a counter-clockwise 45$^\circ$ rotation
creates a NE pointing force on the upper steering element,
and a SW pointing force on the lower steering element,
due to the way these elements have been designed to move
in a concentration gradient.
The coupling of these opposing forces gives a restoring
clockwise torque on the device.
As in Fig.~\ref{mm-ataxis},
only the simplest planar configuration is shown explicitly here;
full three dimensional stability can be obtained
by attaching identical steering elements perpendicular to the page.
To obtain motion (positive or negative chemotaxis),
one can attach a motor with the desired orientation to the central rod.

                \section{Practical Considerations}

An important consideration is the size of the device.
In general terms the larger the lever arm,
the greater the torque, and hence the greater the orientation stability
of the chemitaxis devices described above.
However a larger lever arm also increases the drag,
which reduces the terminal velocity.

The devices have been sketched above
with micro-motors attached to the arms.
Micro-manipulation techniques
in the field of atomic force microscopy
currently allow the routine gluing of colloidal spheres
on the order of 10$\,\mu$m in diameter
to cantilevers on the order of 100$\,\mu$m long
(see, for example, Ref.~\onlinecite{Zhu11} and references therein).
Micro-motors have currently been made
in the form of rods 2$\,\mu$m long and 300$\,$nm in diameter,\cite{Paxton06}
and in the form of 2$\,\mu$m colloidal spheres.\cite{Gibbs10}
Larger motors are likely easier to manufacture,
and such small motors have only been pursued to date
for reasons of buoyancy (see next)
and for the kudos associated with achieving the smallest working micro-motor.

Since the motion of the micro-motors
and of the chemotaxis devices occurs in three-dimensional liquid,
it is important that the effects of gravity be minimised
in order to avoid either sedimentation or flotation of the device
during its motion.
For the 2$\,\mu$m motors mentioned above,
the effects of gravity are small compared to the thermal fluctuations
in the liquid, and  buoyancy effects can be neglected
over experimental time scales.
In the case of the chemotaxis devices proposed here,
micro-manipulation techniques demand somewhat larger micro-motors
and lever arms,
and one may possibly have to consider buoyancy effects.
One solution is to choose connecting rods with positive buoyancy
(e.g.\ polymer),
such that in combination with the negatively buoyant attached motors
the chemotaxis device is overall neutrally buoyant.
One could also possibly attach negatively (e.g.\ tungsten)
or  positively (e.g.\ polymer) buoyant micro-spheres to the
lever arms to achieve neutral buoyancy.

\begin{figure}[t!]
\centerline{ \resizebox{7.cm}{!}{ \includegraphics*{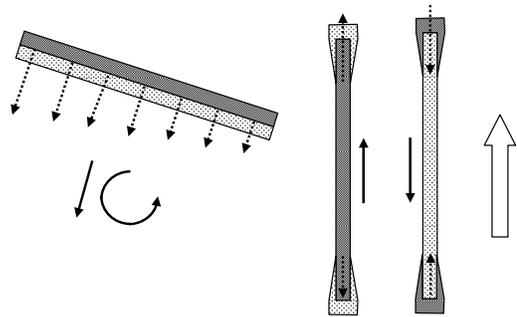} } }
\caption{\label{mm-evap}
Chemotaxis elements made from metal deposition.
The left element is the analogue of the negative chemotaxis device,
Fig.~\ref{mm-ataxis},
and the two right elements correspond
to the respective steering elements in Fig.~\ref{mm-steering}.
The block arrow is the concentration gradient,
the dashed arrows are the forces,
and the solid arrows are the resultant linear and angular motion.
For the steering elements,
a cross section through the rod is shown.}
\end{figure}

Alternatively, it might be possible to avoid the attachment step
and to manufacture a smaller chemotaxis device
by making the lever arms themselves the motors.
This is illustrated in Fig.~\ref{mm-evap},
where three elements for chemotaxis devices are shown.
The elements could be made by metal evaporation and deposition,
as has already been used to manufacture
2$\,\mu$m spherical micromotors.\cite{Gibbs10}
For the case of negative chemotaxis
(left element, Fig.~\ref{mm-evap}),
the element is shown rotated from its stable position
in order to indicate the restoring torque.
It may be seen that it makes no conceptual difference
whether the turning moment arises from a pair of motors
attached at the ends of the lever arm,
or from a continuum of motors distributed along its length.
In the case of the steering elements it is imagined
that the second metal can be deposited on top of the first
at the ends of the rod,
either by appropriate orientation of the rod,
or by the use of shadow masks.
It is not necessary that the entire rod be coated with the first metal.

\begin{figure}[t!]
\centerline{ \resizebox{8.cm}{!}{ \includegraphics*{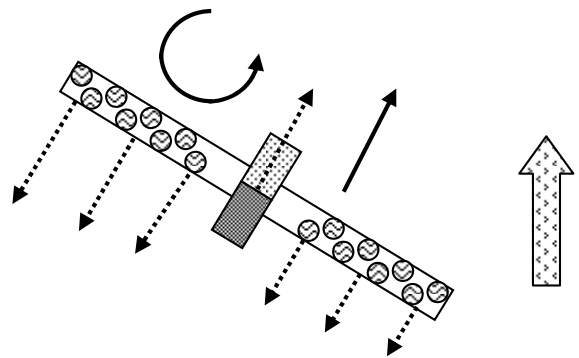} } }
\caption{\label{mm-bio}
Chemotaxis device that separates drive from orientation.
The circles represent binding sites,
the block arrow represents the concentration gradient of the analyte,
the dashed arrows represent the local forces,
and the solid arrows represent the linear and angular motion.}
\end{figure}

Finally,
although the devices proposed above
can be expected to display chemotaxis,
they suffer from a significant limitation,
namely that the chemical used to power the motor is the same
as that used to fix the orientation.
In the real world, organisms that display chemotaxis
respond to a given chemical via its concentration gradient,
and their means of locomotion is quite independent of this stimulus.
The next step in chemotaxis devices would be to separate these functions
so that that motor fuel is different from the chemical gradient
being followed.

Figure~\ref{mm-bio} sketches one possible design for a chemotaxis device
that separates the functions of detection and locomotion.
The idea is that a micro-motor  attached to the center of a rod
moves the device according to the concentration of the chosen propellent
and is insensitive to any concentration gradients.
Along the rod are distributed  binding sites specific for the
analyte of interest.
It is assumed that this analyte  binds reversibly to these sites,
such that the fraction of occupied sites at any time
depends on the local concentration of the analyte.
It is further assumed that bound analytes
increase the drag force on the rod,
which may be reasonable for a polymer or biological macromolecule.
These two design goals are undeniably challenging,
but to the extent that they are satisfied,
and with the addition of an identical arm perpendicular to the page,
the device would have orientation stability,
and it will display positive chemotaxis for the chosen analyte.
Obviously there are many challenges in tuning the parameters
to obtain a working device,
but at least conceptually it illustrates one possible way
that locomotion and steering can be separated
in an artificial chemotaxis device.



\begin{thebibliography}{99}

\bibitem{Paxton05} 
W. F. Paxton, A. Sen, and T. E. Mallouk,
Chem.-Euro.\ J. {\bf 11}, 6462 (2005).


\bibitem{Kline05} 
T. R. Kline, W. F. Paxton,  T. E.Mallouk,  and A. Sen,
Angew.\ Chem.\ Intl Edn {\bf 44}, 744 (2005).

\bibitem{Ozin05}
G. A. Ozin, I. Manners, S. Fournier-Bidoz, and A. Arsenault,
Adv.\ Mater.\ (Weinheim, Ger.) {\bf  17}, 3011 (2005).

\bibitem{Howse07} 
J. R. Howse, R. A. L. Jones, A. J. Ryan, T. Gough,
R. Vafabakhsh, and R. Golestanian,
Phys.\ Rev.\ Let.\ {\bf 99}, 048102 (2007).

\bibitem{Burdick08} 
J. Burdick,  R. Laocharoensuk, P. M. Wheat,  J. D. Posner,  and J. Wang,
J.\ Amer.\ Chem.\ Soc.\ {\bf 130}, 8164  (2008).

\bibitem{Gibbs09} 
J. G. Gibbs and Y. P. Zhao,
Appl.\ Phys.\ Let.\ {\bf 94}, 163104 (2009).

\bibitem{Moran11}
J. L. Moran and J. D. Posner,
J. Fluid Mech. {\bf 680}, 31 (2011).

\bibitem{Zhu11}
L. Zhu, P. Attard, and C. Neto,
Langmuir, {\bf 27},  6712 (2011).

\bibitem{Paxton06}
W. F. Paxton, P. T. Baker, T. R. Kline, Y. Wang, T. E. Mallouk, and A. Sen
J. Am.\ Chem.\ Soc.\ {\bf 128},  14881 (2006).

\bibitem{Gibbs10}
J. G. Gibbs, N. A. Fragnito, and Y. Zhao,
Appl.\ Phys.\ Lett.\ {\bf 97}, 253107 (2010).

\end{thebibliography}
\end{document}